\documentclass[a4paper,twocolumn,
english,aps,pre,floatfix,showpacs]{revtex4}
\usepackage[T1]{fontenc}
\usepackage[latin1]{inputenc}
\usepackage{amsmath}
\usepackage{babel}
\usepackage{graphics}
\usepackage{amssymb}

\begin{document}
 
\title{Dimensional crossover and universal roughness distributions in
Barkhausen noise
}

\author{S.L.A. \surname{de Queiroz}}

\email{sldq@if.ufrj.br}

\affiliation{Instituto de F\'\i sica, Universidade Federal do
Rio de Janeiro, Caixa Postal 68528, 21941-972
Rio de Janeiro RJ, Brazil}

\date{\today}

\begin{abstract}
We investigate the dimensional crossover of scaling properties of 
avalanches (domain-wall jumps) in a single-interface model, used for the 
description of Barkhausen noise in disordered magnets. By varying the 
transverse  aspect ratio $A=L_y/L_x$ of simulated samples, the system 
dimensionality  changes  from two to three. We find that perturbing away 
from $d=2$ is a relevant field. The exponent $\tau$ characterizing the 
power-law scaling of avalanche distributions varies between $1.06(1)$ for
$d=2$ and $1.275(15)$ for $d=3$, according to a crossover function $f(x)$,
$x \equiv (L_x^{-1})^{\phi}/A$, with $\phi=0.95(3)$. We discuss the 
possible relevance of our results to the interpretation of thin-film 
measurements of Barkhausen noise. We also study the
probability distributions of interface roughness, sampled among successive
equilibrium configurations in the Barkhausen noise regime. 
Attempts to fit our data to the class of universality distributions
associated to $1/f^\alpha$ noise give $\alpha \simeq 1-1.1$ for
$d=2$ and $3$ (provided that suitable boundary conditions are used in the
latter case).
\end{abstract}
\maketitle
\section{INTRODUCTION}
\label{intro}
The Barkhausen effect~\cite{barkorig} has long been known in 
magnetism, and reflects the dynamics of domain-wall motion in the 
central part of the hysteresis cycle in ferromagnetic materials.
The intermittent character which is a central feature of Barkhausen
``noise'' (BN) comes to light already in the original experimental setup. 
By wrapping a coil around a sample and ramping an external magnetic field 
at a suitable driving rate, one can detect well-separated voltage pulses 
across the coil, which are induced by sudden changes of magnetic flux. 
These in turn result from the microscopic realignment of groups of 
magnetic moments parallel to the field, i.e.
domain-wall motion. For slow driving rates, the integral of the voltage
amplitude of a given pulse over time is proportional to the change in
sample magnetization, thus giving a measure of the number of spins
overturned in that particular event, or ``avalanche size'', to recall the
terminology frequently used in the study of intermittent phenomena.
Modern experimental techniques allow direct observation of domain-wall
motion via magneto-optical Kerr effect measurements~\cite{pup00,kcs03},
which demands use of a thin-film sample geometry.  

Early proposals for theoretical modelling of BN are reviewed in
Refs.~\onlinecite{abbm,abbm2}, whose authors formulate a Langevin 
description  via Fokker-Planck equations.
More recently, theoretical interest in the description of
the statistical properties of BN has been rekindled, 
as attempts have been made to establish connections with general theories 
of non-equilibrium phase transitions and noise 
phenomena~\cite{cm91,umm95,pds95,czds97,us,tad99,pds99,tad00,tn00,ks00,zcds98,
dz99,dz00,us2} . Here, we shall  be concerned with two such connections. 
The first, motivated by the  thin-film results just alluded to, is the 
crossover between 
universality classes [these latter to be properly defined in the context] 
as one varies the spatial dimensionality of samples; secondly, we
extend recently-developed concepts of universality of
distribution functions for $1/f^\alpha$ noise~\cite{bhp98,bhp00,adgr01,adgr02}
to the scaling properties of domain-wall roughness in BN.

Experimentally, double-logarithmic plots of frequency of avalanche 
occurrence, $P(s)$, against size $s$ turn out to produce unequivocally 
straight sections, $P(s) \sim s^{-\tau}$, often spanning 3-4 orders of 
magnitude, before dropping to zero
for larger sizes~\cite{cm91,bdm94}. Such power-law distribution
of events has been associated to the concepts of self-organized
criticality~\cite{bw90,cm91,obw94,umm95}, 
although other researchers argue that this 
in fact reflects proximity to a standard second-order 
critical point, together with an unusually broad critical region in 
parameter space~\cite{pds95,pds99}. Whatever the interpretation, a 
power-law
decay shows up in assorted models used for numerical simulation of BN,
both those based on the motion of a single interface in a disordered
medium~\cite{abbm,umm95,czds97,us} and those which adopt a picture of
nucleation of multiple domains in a random-field Ising 
system~\cite{pds95}. 

Analogy with the well-established scaling theory of equilibrium phase
transitions suggests that, in this case of a non-equilibrium phenomenon,   
the search for distinct universality classes may lead to a better 
understanding of the basic mechanisms involved. In 
Ref.~\onlinecite{dz00}, experimental measurements of the exponent $\tau$ 
of avalanche distributions for several soft ferromagnetic materials were 
found to separate into two distinct groups, namely $\tau = 1.50 \pm 0.05$
(polycrystalline Fe-Si and partially cristalized amorphous alloys) and 
$\tau = 1.27 \pm 0.03$ (amorphous alloys under stress). It was then
proposed that BN for each group of materials listed above belongs to a 
different universality class of non-equilibrium phase transitions.
While the value of $\tau$ is a fairly plausible indicator of universality,
or lack thereof, between different systems,
many questions (prompted again by analogy with static critical phenomena)
still remain, such as to how many independent exponents there are, and,
of particular interest here, what is the effect of space dimensionality. 
 
Although sample shapes in Ref.~\onlinecite{dz00} were ribbon-like (30 cm 
$\times$ 0.5 cm $\times$ 60 $\mu$m, to quote typical dimensions), this is 
far beyond the thin-film regime, for which thicknesses are of the order 
$5-100$ nm~\cite{pup00,kcs03}.  Thus the behavior reported in
Ref.~\onlinecite{dz00} is expected to 
be characteristic of fully three-dimensional objects. However, when
considering ever thinner samples as in Refs.~\onlinecite{pup00,kcs03}, 
dimensional crossover effects cannot be ruled out from the outset. 

In this work we use a single-interface model, originally introduced in
Ref.~\onlinecite{umm95} for the description of BN. We recall (though
a detailed discussion will be deferred to Section~\ref{conc}) that,
for a fixed space dimensionality $d=2$ or 3, 
numerical values of e.g. the exponent $\tau$ have been found to differ
between nucleation and single-interface models (or even between distinct
formulations of the latter). Here we work under the assumption that
the general features of dimensional crossover to be uncovered are
model-independent, similarly to the presence of power-law avalanche 
distributions. The same assumption is expected to hold as regards
the roughness distributions to be investigated in Section~\ref{sec:4}.

\section{Model and calculational method}
\label{sec:2}
 
Here we shall use the single-interface model introduced in
Ref.~\onlinecite{umm95}. We restrict ourselves
to the adiabatic limit of a very slow driving rate, meaning that
avalanches are regarded as instantaneous (occurring at a fixed value
of the external field). Many experimental setups can be properly
described in this approximation~\cite{umm95,zcds98,us,dz00,kcs03}. 

Simulations are performed on an $L_x \times L_y \times \infty$  geometry,
with the interface motion set along the infinite direction.
The interface at time $t$ is described by its height $h_i \equiv
h(x,y,t)$, where $(x,y)$ is the projection of site $i$ over the 
cross-section. No overhangs are allowed, so $h(x,y,t)$ is 
single-valued. Each element $i$ of the interface experiences a force 
of the form:
\begin{equation}
f_i=u(x,y,h_i)+{k}\left[\sum_{j} h_{\ell_j(i)}- h_i\right]+H_e~,
\label{force}
\end{equation}
where
\begin{equation}
H_e=H-\eta M~.
\label{He}
\end{equation}
The first term on the RHS of (\ref{force}) represents the pinning force,
$u$, and brings quenched disorder into the model by being chosen
randomly, for each lattice site $\vec{r_i} \equiv (x,y,h_i)$,  from a
Gaussian distribution of zero mean and standard deviation $R$. Large 
negative
values of $u$ lead to local elements where the interface will tend to be
pinned, as described in the simulation procedure below.
The second term corresponds to a cooperative interaction among interface
elements, assumed here to be of elastic (surface tension)
type. In this term, $\ell_j(i)$  is the position of the $j$-th nearest
neighbor of site $i$.  
The tendency of this term is to minimize
height differences among interface sites: higher (lower) interface 
elements
experience a negative (positive) force from their neighboring elements.
The force constant $k$ gives the intensity of the elastic coupling,
and is taken here as the unit for $f$. We assume the boundary 
conditions to be periodic along $x$ and free along $y$,
so sites at $y=0$ and $y=L_y$ represent the film's free surfaces and 
have only three neighbors on the $xy$ plane 
(except in the monolayer case $L_y=1$
which is the two-dimensional limit, where all interface sites have two
neighbors).
The last term is the effective driving force, resulting from the applied
uniform external field $H$ and a demagnetizing field which is taken to be
proportional to
$M=(1/L_xL_y)\sum^{L_xL_y}_{i=1} h_i$,
the magnetization  (per site) of the previously flipped spins for a
lattice of transverse area $L_xL_y$.
For actual magnetic samples, the demagnetizing field is not necessarily
uniform along the sample, as implied in the above expression; even when
it is (e.g. for a uniformly magnetized ellipsoid), $\eta$
would depend on the system's aspect ratio. Therefore, our approach amounts
to a simplification, which is nevertheless expected to capture the
essential aspects of the problem. See Ref.~\onlinecite{zcds98} for a
detailed discussion.
Here we use $R=5.0$, $k=1$, $\eta=0.05$, values for which fairly broad
distributions of avalanche sizes and roughness are obtained. 

We start the simulation with a ~flat wall. All spins above it are
unflipped.
The applied field $H$ is set close to
the saturation value of the effective field $H_e$, in order to minimize
transient effects. 
The force $f_i$ is then
calculated for each unflipped site along the interface, and each spin at a
site with $f_i\geq 0$ flips, causing the interface to  move up one step.
The magnetization is updated, and this process continues, with as many
sweeps of the whole lattice as necessary, until
$f_i<0$ for all sites, when the interface comes to a halt.
The external field is then increased by the minimum amount needed to bring
the most weakly pinned  element to motion. The  avalanche size corresponds
to the number of spins flipped between two interface stops.

\section{Scaling of avalanche distributions and dimensional crossover}
\label{sec:3}
We have collected avalanche histograms for varying $L_x$, $L_y$,
in a such a way that the number of interface sites $L_xL_y$ varies between 
800 and 80,000. The aspect ratio $A \equiv L_y/L_x$ was varied between 
essentially zero ($d=2$, one-dimensional interface) and unity ($d=3$,
square interface). For each $L_x$, $L_y$ we generated $10^5$ avalanches.
Although it may take $10^2 - 10^3$ avalanches for a steady-state regime
to be reached (as measured by the stabilization of $H_e$ against external 
field $H$, apart from small fluctuations), we have 
checked that the only distortion introduced by the transient on avalanche 
{\em size} statistics is the one large event occurring at the very start, 
i.e. on departure from the initial, flat-interface, configuration.  
We cannot guarantee this to be so when {\em roughness} is the quantity 
under  investigation, thus data in Section~\ref{sec:4} have been collected 
only under steady-state conditions.

\begin{figure}
{\centering \resizebox*{3.4in}{!}{\includegraphics*{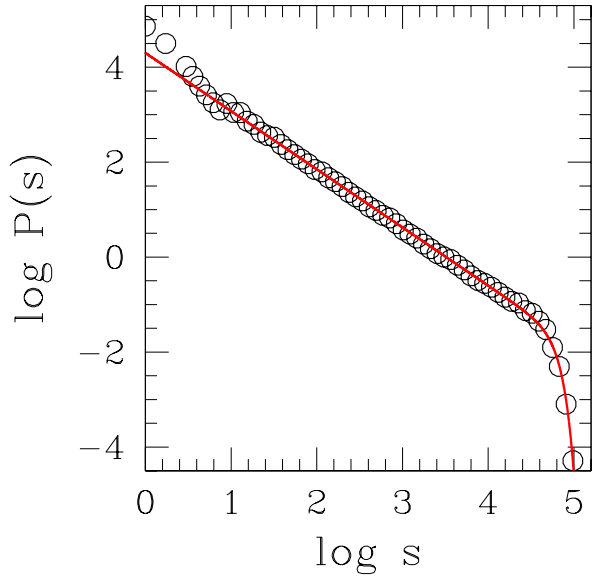}} \par}
\caption{Double-logarithmic plot of avalanche size distribution
for $L_x=60$, $L_y=30$ (circles). Full line is a fit to the form
Eq.~(\protect{\ref{eq:p2(s)}}), for which the optimal parameters are:
$\tau=1.226(6)$, $\delta=3.6(1)$, $s0=5.97(5) \times 10^4$~. }
\label{fig:ps}
\end{figure}

The probability distribution $P(s)$ for avalanche size $s$ is expected
to behave as
\begin{equation}
P(s) = s^{-\tau}\,f \left(\frac{s}{s_0}\right)\quad , 
\label{eq:p(s)}
\end{equation}
where $s_0$ is a cutoff related (in experiment) to domain size and/or
demagnetization effects~\cite{us,dz00}, and
(in simulations) to finite-lattice effects~\cite{umm95,us}, 
or proximity to a critical point~\cite{pds95}. The specific shape 
of the function $f(x)$ has been debated. While a simple exponential
has often been used, either phenomenologically~\cite{bdm94,umm95,us,pds95} 
or (in some special cases) backed by theoretical arguments~\cite{abbm2}, 
Gaussian fits, $f(x) \sim  \exp(-s^2/s_0^2)$~\cite{dz99}, have been 
proposed as well. Going one step further, and at the  same time trying to 
keep the number of fitted parameters to a minimum, here we shall follow 
Refs~\onlinecite{sbms96,tad96} and fit our data to  a {\em stretched} 
exponential:
\begin{equation}
P(s) = A_0\,s^{-\tau}\, e^{-(s/s_0)^\delta}\quad .
\label{eq:p2(s)}
\end{equation}
Apart from the overall normalization factor $A_0$, one then has
three free quantities to fit, which has proved enough for our purposes.
A typical example is displayed in Fig.~\ref{fig:ps}.
For the roughly sixty  $\{L_x,L_y\}$ sets investigated here, the fitted 
value of $\delta$ usually falls in the interval $2.4 - 3.5$, with half a 
dozen cases slightly above that. This is broadly in line with 
$\delta =2.32(6)$ quoted in Ref.~\onlinecite{sbms96}. Though one might
argue that a universal value should hold for this exponent, we feel that
our results are not accurate enough either to prove or disprove such
hypothesis.  

\begin{figure}
{\centering \resizebox*{3.4in}{!}{\includegraphics*{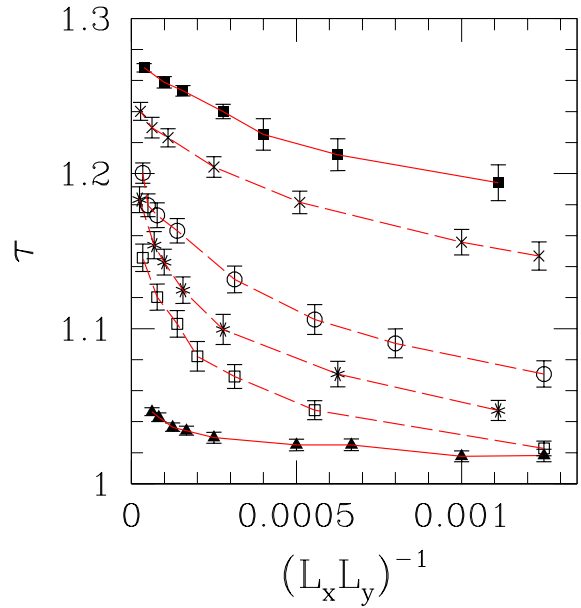}} \par}
\caption{Effective exponent $\tau$ from fits of simulation data to
Eq.~(\protect{\ref{eq:p2(s)}}), against inverse cross-sectional interface
area $(L_xL_y)^{-1}$. Bottom (full triangles linked by full line): 
monolayer, $L_y=1$. Top (Full squares linked by full line): $d=3$,
$L_x=L_y$. Intermediate curves (open symbols connected by dashed
lines): from bottom to top, aspect ratio $A= 0.005$, $0.01$, $0.02$,
$0.1$~.}
\label{fig:exp}
\end{figure}

On the other hand, the exponent $\tau$ which is of central interest
here displays systematic variations both against lattice dimensions
and aspect ratio. These are displayed in Fig.~\ref{fig:exp}. The overall
picture strongly suggests a systematic crossover towards three-dimensional
behavior for any fixed (finite) aspect ratio. In addition to this,
finite-lattice effects are present as well.  

In order to gain a quantitative understanding of this,
we recall general ideas of finite-size~\cite{barber,nig90}
and crossover scaling~\cite{stinch83,cardy96}. 
Crossover phenomena reflect the competition between different
types of (pseudo--)critical behavior in the vicinity of a multicritical
point, at which several characteristic lengths diverge (i.e. their
associated fields approach their respective critical 
values). See e.g. Section VII.A of Ref.~\onlinecite{stinch83} for
an illustration of the well-known case of thermal-geometric crossover in 
dilute magnets near the percolation point. Closer analogy with 
the present case is found in the discussion of dimensional crossover in 
``layer'' magnets~\cite{stinch83,stinch80} (see figure 42 in 
Ref.~\onlinecite{stinch83}). Finite-size scaling can be seen 
as a particular instance of crossover, in which the system's inverse 
finite size $L^{-1}$ is an additional relevant field, driving it away from 
the true criticality which occurs only 
in the thermodynamic limit~\cite{barber,nig90,cardy96}. We now turn to the 
simulational results exhibited in Fig.~\ref{fig:exp}.
Upon increasing sample size, finite-size effects are reduced and the 
trends followed by the respective exponent estimates differ,
depending on whether (i) the aspect ratio $A$ is kept fixed, no matter how 
small its value (all curves except the lower one), or (ii) $L_y =1$ is 
fixed instead (monolayer, lower curve). The latter corresponds to 
ever-decreasing $A$  as $L_x$ increases, and in the $L_x \to \infty$ limit 
is expected to reflect true two-dimensional behavior. 
Considering, for instance, the curve for $A=0.005$, for $L_x$ not very 
large one has exponent estimates closely resembling those of a 
monolayer with a similar cross-sectional ``area''. However, the trend
shown implies that this is in fact an apparent behavior, which arises
only as long as finite-size effects (represented by $L_x^{-1}$) are
more important than the system's three-dimensional character (represented
by a non-zero value of $A$). Increasing $L_x$ while keeping $A$ constant,
the relative importance of these quantities is eventually reversed, and 
one crosses over to three-dimensional behavior. These remarks are 
translated into quantitative statements, as 
follows.   
Since the inverse finite size 
$L_x^{-1}$ and the interface aspect ratio $A$ are both relevant fields,
which drive the system away from true two-dimensional behavior, a
plausible {\em ansatz} for the crossover variable is 
$x = (L_x^{-1})^\phi/A$, where $\phi$ is a crossover exponent, to be 
determined~\cite{stinch83,cardy96}. 
Thus, for any fixed $A \neq 0$, and $L_x \to \infty$ ($x \to 0$), 
three-dimensional features must dominate, while for $x \gg 1$ 
two-dimensional behavior (with finite-size corrections) will take over.

\begin{figure}
{\centering \resizebox*{3.4in}{!}{\includegraphics*{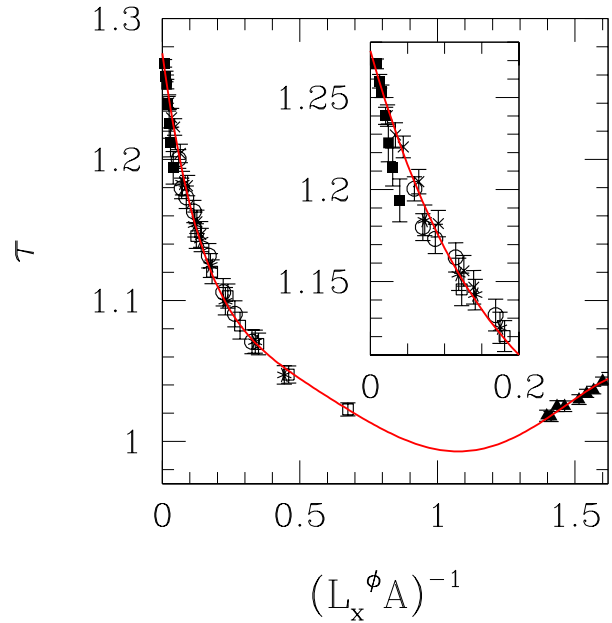}} \par}
\caption{Effective exponent $\tau$ from fits of simulation data to
Eq.~(\protect{\ref{eq:p2(s)}}), against crossover variable
 $x \equiv (L_x^{-1})^\phi/A$, with $\phi=0.95$. Key to symbols is same as 
in caption to
Fig.~\protect{\ref{fig:exp}}~. Full line is a fourth-degree fitting curve
with an exponential tail near $x=0$. Inset shows details of main figure
close to the vertical axis.
}
\label{fig:coll}
\end{figure}

Furthermore, the data of Fig.~\ref{fig:exp} must collapse on the same
curve, when plotted against $x$.
This latter statement gives the operational procedure for determination
of $\phi$. By recalling that the horizontal axis variable in 
Fig.~\ref{fig:exp} can be written as $(L_xL_y)^{-1} = 
x^{2/\phi}\,A^{2/\phi-1}$, and that constant--$A$ curves are further
away from the vertical axis the larger $A$ is, one 
sees that, in order for those curves to collapse one must have $\phi < 2$; 
consideration of the numerical values involved
shows that, in fact, $\phi \lesssim 1$ is needed. 
The choice of $\phi=0.95(3)$,
with the corresponding results exhibited in Fig.~\ref{fig:coll},
reflects the range of $\phi$ for which the collapse of all data for 
non-zero $A$ (onto the left-hand side of the diagram) is visually  deemed 
to be best. The fact that the $d=2$ data are segregated towards the 
right is to be expected in this context, as they belong to a different 
side of the crossover ($x>1$) where three-dimensional effects vanish.

The intersection of the scaling  curve with the vertical
axis provides an estimate of the three-dimensional scaling exponent.
The result is $\tau (d=3)= 1.27(1)$. An {\em ad hoc} parabolic fit
against $1/L_x$, including only data for $A=1$ and $L_x \leq 160$, 
gives $\tau (d=3)=  1.28(1)$, where the largest contribution to the 
uncertainty comes from spread between extrapolations that either do or do 
not include small-lattice data ($L_x=30$, $40$). 
As we have seven lattice sizes available for $A=1$, use of 
parabolic fits already gives us at most four degrees of freedom.
Given the limited range of data available, attempting to improve 
estimates by including higher-order corrections would therefore not seem 
justifiable.  
Since neither extrapolation procedure
appears to be obviously superior to the other, our final quote encompasses
both results: $\tau(d=3)=1.275(15)$.

In Ref.~\onlinecite{umm95},
$\tau(d=3)=1.13(2)$ is quoted for $L_x=40$, while our corresponding
result is $1.21(1)$. 
This difference arises mainly from distinct 
fitting procedures, in particular those authors' apparent use of a fixed 
$\delta=1$ for the cutoff. Indeed, by keeping $\delta=1$, the best fit 
of our data is for $\tau(d=3)=1.12(4)$, though at the cost of 
increasing the $\chi^2$ per degree of freedom by one order of 
magnitude compared with the variable--$\delta$ fitting scheme. 

A similar parabolic fit of the two-dimensional data 
gathered on the right-hand corner of Fig.~\ref{fig:coll}
(using $4000 \leq L_x \leq 16000$), gives $\tau(d=2)=1.06(1)$ .
This is broadly in accord with the result of Ref.~\onlinecite{umm95}, 
namely $\tau(d=2)=1.00(1)$ (using $500 \leq L_x \leq 5000$).

\section{Roughness distributions and $1/f^\alpha$ noise}
\label{sec:4}
In this section, we consider only two-dimensional ($L_x=1$) and fully
three-dimensional ($L_x=L_y$) systems.
We have collected data on interface roughness, in order to analyze them in 
the context of universal fluctuation distributions. 
An important reason for interest in universal distributions is that they 
have no adjustable parameters~\cite{bhp98,bhp00,adgr01,adgr02}. The fact 
that a given  property of a system behaves according to one of such 
distributions
is then expected to indicate the universality class to which it belongs.
For non-equilibrium problems such as is the case here, the task of
connecting exponents (or features of distributions) to universality 
classes is far from accomplished (see e.g. the discussion in 
Ref.~\onlinecite{adgr02}). Accordingly, our purpose here is simply to
identify what universal distribution (if any) is followed by
the scaled interface roughness for the current model.

In order to make contact with previous work on interface roughness at
depinning transitions, we recall some basic ideas.

At the end of each avalanche, we measured the roughness ${\cal R}$ of the 
instantaneous interface configuration at time $t$ in the usual way, as 
the rms fluctuation of the interface height: 
\begin{equation}
{\cal R}(t) 
=\left[\left(L_xL_y\right)^{-1}\,\sum_{i=1}^{L_xL_y}\left(h_i(t)-\overline{h}(t)
\right)^2\right]^{1/2}\ ,
\label{eq:rough1}
\end{equation}
where $\overline{h}(t)$ is the average interface height at $t$. As the
avalanches progress, one gets a sampling of successive equilibrium
configurations; the ensemble of such configurations yields a distribution 
of the relative frequency of occurrence of ${\cal R}$. 
In order to get clean distributions, we 
have seen that the number of events considered must be ${\cal O}(10^6)$,
i.e. one order of magnitude larger than the samples used for the analysis
of size distributions in Section~\ref{sec:3}. We have used only
steady-state data. i.e. after the stabilization of $H_e$ of 
Eq.~(\ref{He}) against external  field $H$. This way, our implicit 
assumption that successive interface configurations are stochastically
independent gains plausibility. Similar ideas were invoked in
Ref.~\onlinecite{forwz94} to justify a mapping of the steady state of a 
deposition-evaporation model onto a random-walk problem. 

The roughness exponent $\zeta$ is related to the finite-size scaling 
of the first moment of the distribution~\cite{bar95}:
\begin{equation}
\langle{\cal R}_L\rangle \sim L^{\zeta}\ \ ,
\label{eq:zeta}
\end{equation}
where the angular brackets stand for averages over the ensemble of
successive interface configurations of an interface with transverse 
dimension $L$. We have estimated $\zeta$ from 
power-law fits to our data. For $d=2$ (one-dimensional interface)
using $400 \leq L_x \leq 1200$ we get $\zeta (2d)= 1.24(1)$, which
compares  well with the usually accepted $\zeta \simeq 1.25$ for the
quenched Edwards-Wilkinson universality class~\cite{les93,ma95,mbls98,rhk03}. 
For $d=3$ (two-dimensional interface)
we used $30 \leq L_x=L_y \leq 80$ and two alternative sets of boundary 
conditions, namely mixed (MBC) i.e. free along $x$ and periodic along $y$,  
as described in Section~\ref{sec:2}, and periodic along both $x$ and $y$ (PBC).
The results are, respectively, $\zeta(d=3,{\rm MBC}) = 0.87(1)$ and
$\zeta(d=3,{\rm PBC}) = 0.71(1)$. While the latter value 
is not far from $\zeta \simeq 0.75$ for the corresponding quenched 
Edwards-Wilkinson  model~\cite{les93,rhk03}, the former seems  
difficult to relate to existing results. 

At this point one might invoke universality ideas and claim that the
difference between three-dimensional estimates must be a finite-size
effect. Investigating this directly, e.g. by performing simulations for
larger $L$, would be straightforward but time-consuming. However, as we
shall see in what follows, this issue can be addressed more efficiently by
returning to our main goal in this Section, i.e. by looking at the full
distributions, rather than examining only selected moments (as is the case
of the scaling for extraction of $\zeta$). This is because mounting
evidence indicates that, in general, width distributions  decouple
into the product of a single size-dependent scale and a universal 
(size-independent) scaling function~\cite{forwz94,adgr01,adgr02,rkdvw03},
that is
\begin{equation}
P({\cal R}) = (1/\sigma)\,\Phi({\cal R}/\sigma)\ ,\quad 
\sigma^2=\langle({\cal R} -\langle{\cal R}\rangle)^2\rangle\ .
\label{eq:pvsphi}
\end{equation}
Therefore, if the functional form  $\Phi(z)$ varies depending e.g.
on whether mixed or periodic boundary conditions are used,
one is on safer grounds to assume that this reflects differences in the 
respective universality class. In this context, it must be recalled
that dependence of scaling quantities on boundary conditions is an 
often--encountered feature when dealing with 
fluctuation phenomena~\cite{adgr01,adgr02}. One illustration of this,
which is of great relevance here,
is that one of the requirements for a class of universal 
(time) fluctuation distributions to hold is that they be  periodic in 
time~\cite{adgr01,adgr02,forwz94}.

In our case, this feature is replaced by periodic boundary conditions {\em 
in space} (the same reasoning was used for the deposition-evaporation 
model of Ref.~\onlinecite{forwz94}). Strictly speaking, this time-space
correspondence is only true for the one-dimensional interfaces of the
two-dimensional version of our model. However, the 
mixed  boundary conditions (MBC) used in Sec.~\ref{sec:3}  
suggest that an extension of our investigation to the three-dimensional
case may not be unjustified. As regards fully periodic boundary 
conditions (PBC), even though it is not obvious that the analogy can 
be pushed that far, we decided to analyse the respective data for 
completeness.  

We analysed the roughness distributions for $d=2$ and for the fully
three-dimensional case (i.e. aspect ratio $L_y/L_x=1$), the latter 
both with MBC and PBC. As shown below, finite
transverse dimensions are of negligible import as far as the 
scaling functions $\Phi(z)$ are concerned, thus giving further support
to the assumption that Eq.~(\ref{eq:pvsphi}) holds.

We have compared our results against the family of roughness distributions 
for periodic $1/f^\alpha$ noise, described in Ref.~\onlinecite{adgr02}. 
The roughness of a time signal, and its connections via Fourier 
transform with the frequency spectrum (and thus with the respective 
$\alpha$), are described at length in Section II of that reference.
The idea of  fitting spatial roughness data to the roughness of time
signals (e.g. random-walks) is 
well-known in the study of interface fluctuations~\cite{bar95}.
The new feature here is that we have at our disposal a family of
distributions whose shape varies smoothly against the single basic
parameter $\alpha$ (but, apart from that, are strictly 
parameter-independent). We expect that the value of  $\alpha$  
which
best fits our data should be connected to the underlying universality
class of thew avalanche model used here. We recall 
that ours is a phenomenological study, since at present very little
is known regarding the (physical) causal relationships between fluctuation
distributions and universality classes for out-of-equilibrium 
transitions~\cite{bhp98,bhp00,adgr01,adgr02}. 
  
The quantity to be compared against a universal form is the distribution 
of the deviation from the average, scaled by the standard 
deviation (this is termed {\em scaling by the variance} in 
Ref.~\onlinecite{adgr02}):
\begin{equation}
y \equiv \frac{{\cal R} -\langle{\cal R}\rangle}{\sigma}\ .
\label{eq:scalevar}
\end{equation}

\begin{figure}
{\centering \resizebox*{3.4in}{!}{\includegraphics*{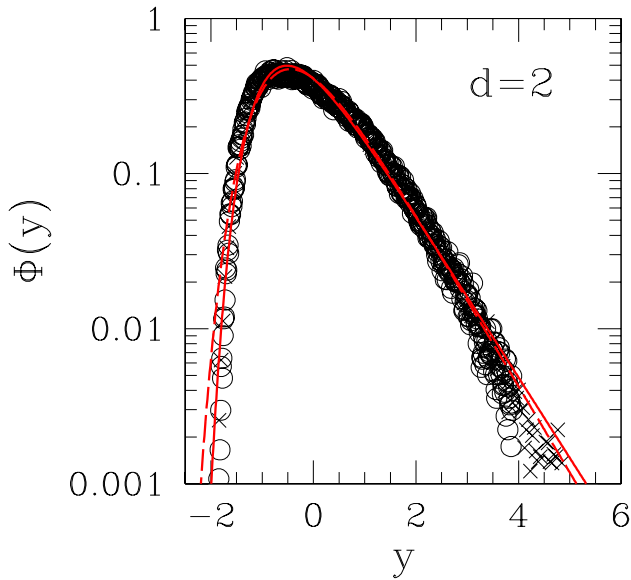}} \par}
\caption{Scaled probability distributions $\Phi(y)$ in $d=2$ for $y$ 
defined in
Eq.~(\protect{\ref{eq:scalevar}}). Crosses: $L_x=400$;
circles: $L_x=1000$. Lines are roughness 
distributions for $1/f^\alpha$ noise given in 
Ref.~\protect{\onlinecite{adgr02}}~, with $\alpha=1.15$ (full) and $1$
(dashed). 
}
\label{fig:rdist2d}
\end{figure}

\begin{figure}
{\centering \resizebox*{3.4in}{!}{\includegraphics*{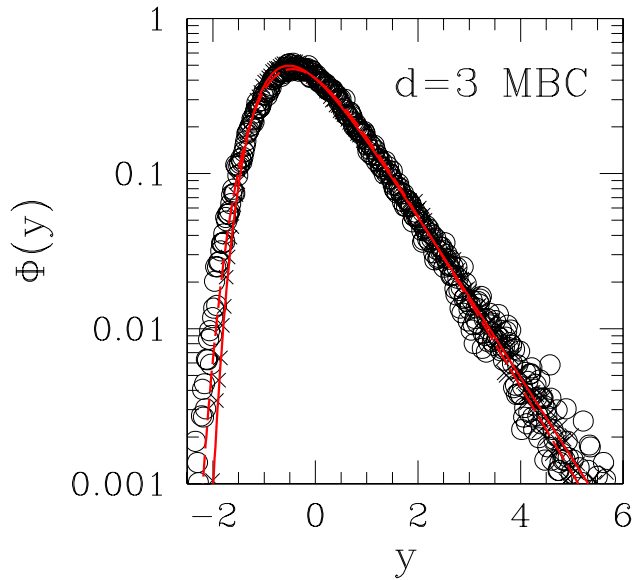}} \par}
\caption{Scaled probability distributions $\Phi(y)$ in $d=3$ with MBC, for 
$y$ defined in
Eq.~(\protect{\ref{eq:scalevar}}). Crosses: $L=30$;
circles: $L=80$. Lines are roughness 
distributions for $1/f^\alpha$ noise given in 
Ref.~\protect{\onlinecite{adgr02}}~, with $\alpha=1.15$ (full) and $1$
(dashed). 
}
\label{fig:rdist3d}
\end{figure}

\begin{figure}
{\centering \resizebox*{3.4in}{!}{\includegraphics*{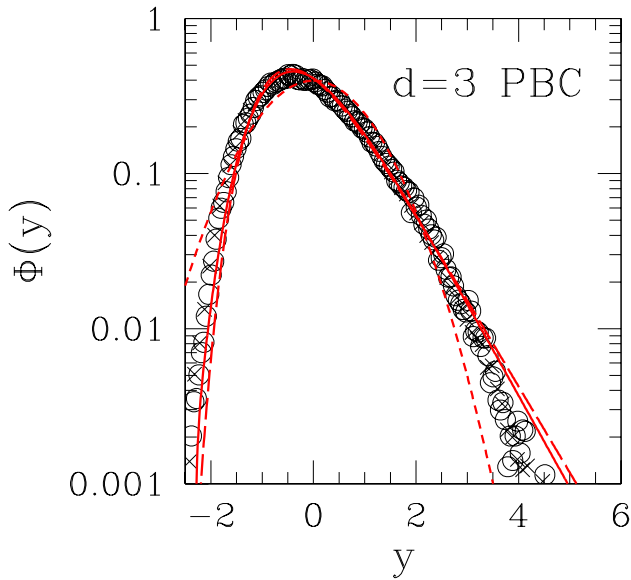}} \par}
\caption{Scaled probability distributions $\Phi(y)$ in $d=3$ with PBC for 
$y$ defined in
Eq.~(\protect{\ref{eq:scalevar}}). Crosses: $L=30$;
circles: $L=80$. Lines are roughness 
distributions for $1/f^\alpha$ noise given in 
Ref.~\protect{\onlinecite{adgr02}}~, with $\alpha=0.9$ (full), $1$
(dashed) and $0.5$ (short-dashed). 
}
\label{fig:rdist3dp}
\end{figure}

\begin{figure}
{\centering \resizebox*{3.4in}{!}{\includegraphics*{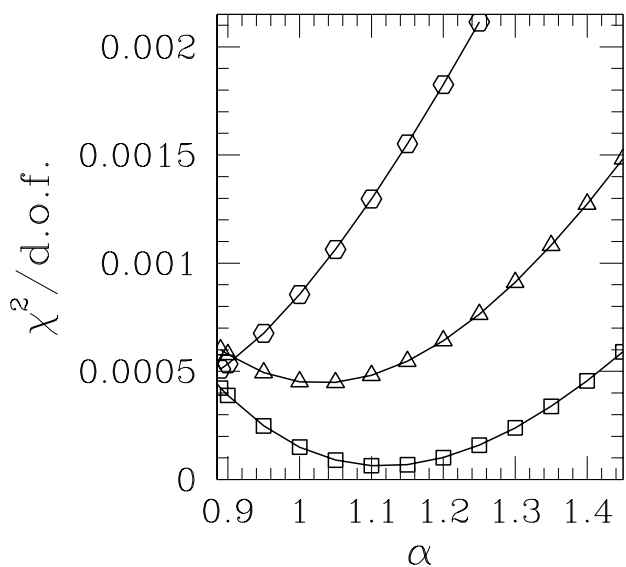}} \par}
\caption{ $\chi^2$ per degree of freedom ($\chi^2/{\rm d.o.f.}$)
for fits of simulation data to analytical forms of $1/f^\alpha$
distributions, against $\alpha$. Triangles: $d=2$, $L_x=400$;
Squares: $d=3$ MBC, $L=40$; hexagons, $d=3$ PBC, $L=40$.
}
\label{fig:chi2}
\end{figure}

Our data are depicted respectively in Fig.~\ref{fig:rdist2d} for $d=2$,
Fig.~\ref{fig:rdist3d} for $d=3$ with MBC, and Fig.~\ref{fig:rdist3dp} for 
$d=3$ with PBC. In each figure, we have plotted data corresponding 
to only two values of $L$ (``small'' and ``large''), in order to
illustrate that the $L$--dependence of the scaled distribution is
negligible, without cluttering the diagram with intermediate--$L$ results. 
In all figures, the analytic curve corresponding to $\alpha=1$ is shown
as a dashed line, in order to ease comparison between different cases.
In general, the scatter of simulational data at the low-roughness end of
the distribution is much smaller than that at the high end, thus we
might give more weight to the former region when judging the quality
of fit. 
Using this criterion, one sees that the $d=2$ data appear to be closer to 
the $\alpha=1.15$ curve than to
that for $\alpha=1$, while the situation is reversed in the $d=3$ case
with MBC. Furthermore, for $d=3$ with PBC the best fit is undoubtedly
for $\alpha < 1$. For this latter case, we also display a Gaussian
distribution, which corresponds to $\alpha=0.5$~\cite{adgr02}. 
We have tried to quantify the above remarks, by investigating the behavior
of the $\chi^2$ per degree of freedom ($\chi^2/{\rm d.o.f.}$)
for fits of our data to the analytical distributions, against varying 
$\alpha$. We have included only data for which $\Phi(y) \ge 10^{-3}$,
i.e. those displayed in Figs.~\ref{fig:rdist2d},~\ref{fig:rdist3d}, 
and~\ref{fig:rdist3dp}. Though, in principle, the distributions can
be evaluated in closed form~\cite{adgr02}, we ran into serious
numerical problems for $\alpha \lesssim 0.9$ in the region $y \lesssim 
-1$. Fortunately, as shown in Fig.~\ref{fig:chi2}, this does not matter 
much as long as the $d=2$ and $d=3$ (MBC) cases are concerned, because
the respective $\chi^2/{\rm d.o.f.}$ clearly exhibit minima located
slightly above $\alpha=1$. On the other hand, this means that for  $d=3$ 
with PBC we were not able to follow the trend shown in the Figure into the
region $\alpha < 0.9$, where it is clear that a minimum of the
corresponding $\chi^2/{\rm d.o.f.}$ must be located.   
On the other hand, the  Gaussian ($\alpha=0.5$) distribution
shown in Fig.~\ref{fig:rdist3dp} evidently overshoots the desired
corrections, so 
one can be sure that the best fit will be in the interval
$0.5 < \alpha < 0.9$.

\section{Discussion and Conclusions}
\label{conc}

The data displayed in Section~\ref{sec:3} show that, in the interface
model introduced in Ref.~\onlinecite{umm95}, the power-law behavior of
avalanche statistics is characterized by an effective exponent, which
varies continuously both with the interface's linear dimensions and aspect
ratio. By means of a finite-size and crossover analysis, we have
demonstrated that such continuous variation in fact reflects crossover 
towards three-dimensional behavior, for any non-zero aspect ratio.

The implications of this for the interpretation of experimental BN results
in thin films must be worked out carefully. In Ref.~\onlinecite{pup00},
the value $\tau=1.1$ is given for Fe films. The authors of
Ref.~\onlinecite{kcs03} quote $\tau \simeq 1.33$ for Co
films and conclude that their setup is a two-dimensional realization of 
the (single-interface) model of Ref.~\onlinecite{czds97}, for which
$\tau=4/3$ in $d=2$ and $3/2$ in $d=3$~\cite{zcds98,vsc00}.
On the other hand, as seen above, the single-interface model considered 
here gives $\tau=1.06(1)$ in $d=2$ and $1.275(15)$ in $d=3$. Thus one 
might interpret both sets of experimental results as reflecting a 
crossover towards three-dimensional behavior. As regards the specific 
features of the experimental investigations, one must ask: (i) which,
if any, of the two models applies to the corresponding microscopic 
description,  and (ii) how far along, quantitatively, is the 
dimensional crossover for the thin-film geometries used.  

While the visual evidence displayed in Ref.~\onlinecite{kcs03} is 
convincing proof that a single-interface picture applies in that case,
for a definite answer to (i) one must look at the differences between
the models. As far as the power-law scaling of avalanche distributions
is concerned, the model of Ref.~\onlinecite{czds97} differs from the one 
considered here by the introduction of a non-local kernel due to dipolar 
interactions. The value $\tau=4/3$ quoted in Ref.~\onlinecite{kcs03} 
relies on assuming that the form taken by this kernel in momentum space is 
$\sim q^\mu$, ($q=$ wavevector) with $\mu=1$~\cite{vsc00}. In $d=3$ 
the same theory gives $\tau=3/2$ for $\mu=1$, and $\tau =5/4$ for 
$\mu=2$~\cite{dz00}. Both values have been found to good 
approximation in experiments on fully three-dimensional systems (thus 
defining distinct universality classes), as  recalled in the 
Introduction~\cite{dz00}. For the thin-film cases it is not clear, 
without a detailed analysis of the specific materials involved, whether 
the non-local kernel is of sufficient import to drive avalanche scaling 
towards the $d=2$ behavior predicted for the model of 
Ref.~\onlinecite{czds97}.

Turning to question (ii), recall that the evolution of an
interface along a $400 \times 320$ $\mu$m$^2$ area of a 25-nm film
is shown in Ref.~\onlinecite{kcs03}. Translating to the language of 
Section ~\ref{sec:3}, this would  correspond to a transverse aspect ratio 
$A= 25 {\rm nm}/320 \mu{\rm m} \simeq 8 \times 10^{-5}$ 
(this in an upper bound, as the film's transverse dimensions are likely to 
be larger than the area shown). Though we do not think that the model results  
depicted in Fig.~\ref{fig:exp} are quantitatively accurate enough, one 
must keep in mind the possibility that the effective experimental behavior 
still is very close to the two-dimensional limit. Indeed, the 
simulational curve for $A=5 \times 10^{-3}$ already shows a value of 
$\tau$ rather close to that for the two-dimensional case, along an 
extended portion of the Figure. The surest way 
to settle this matter would be by performing a series of experiments on
films of the same composition and varying thicknesses, in order to
produce a full picture of the dimensional crossover. We hope 
experimentalists will be motivated by the present results.

As regards the search for universal roughness distributions in 
Section~\ref{sec:4}, for now we quote (from Figure~\ref{fig:chi2} and
the associated remarks) $\alpha \simeq 1.05$ ($d=2$); $\alpha \simeq 1.15$ 
($d=3$, MBC) and $0.5 < \alpha < 0.9$ ($d=3$, PBC).
It thus appears that the boundary conditions do have significant influence
in this context, a fact which remains to be more fully understood.
Although the search for the physical origins of $1/f$ 
noise~\cite{weiss88} is clearly of great interest, it appears that, at 
least for non-equilibrium phenomena as is the case here, we are still at a 
very preliminary stage. Again, it is hoped that the present results will
motivate further research. Measurements of the roughness distributions for
alternative models of BN~\cite{pds95,czds97} would be a natural extension
of the this work, in order to check whether the above-quoted values 
of $\alpha$ are indeed universal within this subset of avalanche models. 
 
\begin{acknowledgments}

The author thanks Belita Koiller and Monica Bahiana for many interesting 
discussions and suggestions. The
research of S.L.A.d.Q. was partially supported by the Brazilian agencies
CNPq (Grants No. 30.1692/81.5 and No. 47.4715/01.9), FAPERJ (Grant
No. E26--152.195/2002), FUJB-UFRJ and Instituto do Mil\^enio de
Nanoci\^encias--CNPq.
\end{acknowledgments}

\bibliography{biblio}  
\end{document}